\documentclass{article}

\usepackage{aaai}
\usepackage{times}

\usepackage{latexsym}
\usepackage{amsmath,amsthm,amssymb,stmaryrd, bm}

\usepackage{url}

\usepackage{xcolor}
\usepackage{tikz}

\usepackage{listings}
\newcommand{\ind}{\mathsf{ind}}
\newcommand{\Mmf}{\mathfrak{M}}
\newcommand{\avec}[1]{\boldsymbol{#1}}
\renewcommand{\int}{\mathsf{int}}

\newcommand{\omq}{\boldsymbol Q}
\newcommand{\D}{\mathcal{D}}
\renewcommand{\O}{\mathcal{O}}
\newcommand{\diamondplus}{%
  \raisebox{-.15ex}{\begin{tikzpicture}
    \useasboundingbox (-0.7ex, -0.9ex) rectangle (0.7ex, 0.9ex);
    \node (w) at (-0.8ex,0) {};
    \node (e) at (+0.8ex,0) {};
    \node (s) at (0,-.9ex) {};
    \node (n) at (0,+.9ex) {};
    \draw (n.center) -- (e.center) -- (s.center) -- (w.center) -- (n.center);
    \draw (n.center) -- (s.center);
    \draw (e.center) -- (w.center);
  \end{tikzpicture}}}
\newcommand{\diamondminus}{%
  \raisebox{-.15ex}{\begin{tikzpicture}
    \useasboundingbox (-0.7ex, -0.9ex) rectangle (0.7ex, 0.9ex);
    \node (w) at (-0.8ex,0) {};
    \node (e) at (+0.8ex,0) {};
    \node (s) at (0,-.9ex) {};
    \node (n) at (0,+.9ex) {};
    \draw (n.center) -- (e.center) -- (s.center) -- (w.center) -- (n.center);
    \draw (e.center) -- (w.center);
  \end{tikzpicture}}}

\newcommand{\hMTL}{\textsl{HornMTL}}
\newcommand{\dMTL}{\textsl{datalogMTL}}

\begin{document}
%
\title{Metric Temporal Logic for Ontology-Based Data Access over Log Data}
\author{
\begin{tabular}{ccc}
D.~Calvanese,\ E.~G\"uzel Kalayc{\i},\ V.~Ryzhikov,\ G.~Xiao &\and & M.~Zakharyaschev\\ \small
Faculty of Computer Science && \small Department of Computer Science\\%
\small Free University of Bozen-Bolzano, Italy && \small Birkbeck, University of London, UK
\\
\small \{calvanese,kalayci,ryzhikov,xiao\}@inf.unibz.it && \small michael@dcs.bbk.ac.uk
\end{tabular}
}

\nocopyright
\maketitle
\begin{abstract}
 We present a new metric temporal logic $\hMTL$ over dense time and its datalog extension $\dMTL$. The use of $\dMTL$ is demonstrated in the context of ontology-based data access over meteorological data. We show decidability of answering  ontology-mediated queries for a practically relevant non-recursive fragment of $\dMTL$. Finally, we discuss directions of the future work, including the potential use-cases in analyzing log data of engines and devices.
\end{abstract}


\section{Introduction}

The aim of ontology-based data access (OBDA)~\cite{PLCD*08} is, on one hand, to represent the information from various heterogeneous data sources in a unified and conceptually transparent way by means of \emph{mappings}. On the other hand, the \emph{ontology language} allows one to define concepts in terms of other concepts, and thereby represent frequently used query patterns as reusable concepts. The end-user, in that case, can obtain the required information by means of simple conceptual queries and is not required to know neither the structure of the source data nor the definitions of the concepts he is using.

Due to up-to-date requirements of industry (see, e.g., \cite{DBLP:conf/semweb/KharlamovSOZHLRSW14}) the OBDA approach is  being actively adopted in the context of the temporal data of streams and logs. Initially, only the classical non-temporal ontology languages were considered to mediate the access to temporal data~\cite{DBLP:conf/rr/Gutierrez-BasultoK12,OMNZK:13,BBL13,DBLP:conf/rr/KlarmanM14}. Later, the ontology languages with temporalized concepts were studied in this context~\cite{DBLP:conf/ijcai/ArtaleKKRWZ15,IJCAI1toappear,GJK-IJCAI16}. Such concepts are defined by means of linear temporal logic (LTL); for example, the axiom
\begin{multline*}
{\sf Hurricane} \leftarrow \mathsf{HurricaneForceWind} \land{} \\
 \mathbf{X}^-\mathsf{HurricaneForceWind}
\end{multline*}
defines a hurricane as hurricane force wind lasting for 1 hour ($\mathbf{X}^-$ is the \emph{previous time} LTL operator). One easily notices that this definition works only if the temporal data arrives strictly in hourly periods, such as {\sf 13:21}, {\sf 14:21}, etc. If these periods are smaller and have a fixed length, the definition above can still be adjusted by using the conjunction of the form ${\sf HFW} \land \mathbf{X}^-{\sf HFW} \land \mathbf{X}^- \mathbf{X}^- {\sf HFW} \land \dots$. However, first, having the data with fixed-period timestamps is not always a realistic assumption, and, second, doing the adjustment above contradicts the OBDA philosophy, where the ontology user is not required to have knowledge of the structure of the data sources. Therefore, the following definition would be more natural
$$\mathsf{Hurricane} \leftarrow  \boxminus^{\leqslant 1h}_{\geqslant 0} \mathsf{HurricaneForceWind},$$
where $\boxminus^{\leqslant 1h}_{\geqslant 0}$ is a metric temporal operator \emph{during the previous hour}. The logic required to express such statements is a kind of \emph{metric temporal logic} or \emph{modal logic of metric spaces}; see~\cite{Koymans1990,DBLP:conf/birthday/KuruczWZ05} for surveys and further references.

In this paper, we introduce a metric temporal logic $\hMTL$ with the operator $\boxminus^{\lhd d}_{\rhd e}$, where $\rhd$ is either $>$ or $\geqslant$ (and similarly for $\lhd$) and $e,d$ are time \emph{distances}, its future analogue $\boxplus^{\lhd d}_{\rhd e}$, as well as their duals $\diamondplus^{\lhd d}_{\rhd e}$ and $\diamondminus^{\lhd d}_{\rhd e}$. We interpret this logic over a \emph{dense} temporal domain. The reason for not considering a \emph{discrete} domain is that we want to abstract from the \emph{granularities} of time (periods of timestamps) in the data sources. In our logic, we allow the statements of the form $P@\iota$, where $\iota$ is an interval specified by a pair of time instants, to represent the conceptualized temporal data. The meaning of, say, $P@(t_1, t_2]$ is that $P$ holds at all times $t$ between $t_1$ (not including it) and $t_2$ (including it). We assume that we can convert data from any source with timestamped tuples to this format by means of mappings. For example, if a source contains the information of temperature measurements taken every hour, such as {\sf 13:21: -1$^\circ$C}, {\sf 14:21:  2$^\circ$C}, {\sf 15:21:  -1$^\circ$C}, etc., we can conceptualize them as the statements {\sf $\sf PositiveTemp@$(13:21, 14:21]}, etc. Note that whether to include the ends of intervals or not, as well as whether to consider 2$^\circ$C to be the case in the hour preceding or following {\sf 14:21}, is the choice of the mapping designer. We then extend $\hMTL$ to $\dMTL$ that also allows for standard Datalog reasoning about objects of the application domain (weather stations, cities, sensors, etc.).

We present a few preliminary results on $\dMTL$. First, we describe a use-case of OBDA over meteorological data with SQL mappings to a large real-world weather database and $\dMTL$ as an ontology language. Second, we develop an ontology-mediated query answering algorithm for a non-recursive fragment $\dMTL^\Box_{\it nr}$ of $\dMTL$. Finally, we report some preliminary evaluation results showing the feasibility of our approach.

\section{$\hMTL$ and $\dMTL$}

\paragraph{Syntax.} We consider a propositional temporal logic $\hMTL$ with the set of propositional variables $P_0, P_1,
\dots$ over the temporal domain $\mathfrak T$ isomorphic to $(\mathbb R, \leqslant)$ with $0$ and arithmetic operations $+,-$.  That is, we assume \emph{dense} time. Let $\int(\mathfrak{T})$ be the set of (non-empty) \emph{intervals on} $\mathfrak{T}$, which are of the form $[t_1, t_2]$, $[t_1, t_2)$, $(t_1, t_2]$, and $(t_1, t_2)$, where $t_i \in \mathfrak{T} \cup  \{-\infty, \infty\}$, $\langle$ is either $($ or $[$, and $\rangle$ is either $)$ or $]$. (We do not distinguish between the intervals $\langle t_1, \infty]$ and $\langle t_1, \infty)$, consider $\langle \infty, \infty \rangle$ to be empty, and analogously for $-\infty$. We also assume that $\leq$ is defined on $\mathfrak{T} \cup \{-\infty, \infty\}$ and $+,-$ are defined on pairs of elements from $\mathfrak T$ and $\{-\infty, \infty\}$, in a standard way.) Define a \emph{data instance} $\D$ as a non-empty finite set of \emph{data assertions} (or \emph{facts}) of the form:
$$P_i@\iota,$$
where $P_i$ is a propositional variable and $\iota \in \int(\mathfrak{T})$. 

We use the temporal operators of the form:
\begin{itemize}
 \item[--] $\boxplus^{\lhd d}_{\rhd e}$ (always between $e$ and $d$ in the future),
 \item[--] $\boxminus^{\lhd d}_{\rhd e}$ (always between $e$ and $d$ in the past),
 \item[--] $\diamondplus^{\lhd d}_{\rhd e}$ (sometime between $e$ and $d$ in the future),
 \item[--] $\diamondminus^{\lhd d}_{\rhd e}$ (sometime between $e$ and $d$ in the past),
\end{itemize}
 where $\lhd$ is either $<$ or $\leqslant$, $e,d$ are \emph{distances}, that is, \emph{positive} elements of $\mathfrak{T}$, and $\rhd$ is either $>$ or $\geqslant$. Thus, e.g., $\boxplus^{< d}_{\geqslant e}$ expresses `always between $e$ and $d$ in the future including $e$ and excluding $d$' and similarly for the other operators. We also impose the following consistency requirement on every operator $\mathbf{O}^{\lhd d}_{\rhd e}$ (henceforth we assume $\mathbf{O} \in \{ \boxplus, \boxminus, \diamondplus, \diamondminus \}$, $\mathbf{\Box} \in \{ \boxplus, \boxminus\}$, and $\Diamond \in \{\diamondplus, \diamondminus \}$):
\begin{itemize}
  \item[--] there exists $t \in \mathfrak{T}$ such that $t \rhd e$ and $t \lhd d$.
\end{itemize}
Propositional \emph{literals} are defined by the following grammar:
$$\lambda \ ::= \ \ P_i \ \mid \ \ \mathbf{O}^{\lhd d}_{\rhd e} \lambda.
$$
An \emph{ontology}, $\O$, is a finite set of \emph{axioms} of the form:
\begin{equation}\label{clauses}
  \lambda \leftarrow \lambda_1 \land \dots \land \lambda_k, \qquad \bot \leftarrow \lambda_1 \land \dots \land \lambda_k.
\end{equation}
A \emph{knowledge base} (KB) is a pair $(\O, \D)$.
\paragraph{Semantics.} Consider an interpretation $\Mmf = (\mathfrak{T}, \cdot^\Mmf)$ such that $P_i^\Mmf \subseteq \mathfrak{T}$ for each propositional variable $P_i$ and write $\Mmf, t \models P_i$ when $t \in P_i^\Mmf$ for $t \in \mathfrak{T}$. As usual, it is assumed that $\Mmf, t \not \models \bot$ for all $t \in \mathfrak{T}$. We extend the definition of $\models$ to $\lambda$ as follows:
\begin{align}\label{prop-semantics1}
\notag \Mmf, t \models \boxplus^{\lhd d}_{\rhd e}\lambda \quad &\text{iff} & \Mmf, t' \models \lambda \text{ for all } t' \text{ such that }\\
 & & t' -t \rhd e \text{ and }t' - t \lhd d,\\
\notag \Mmf, t \models \boxminus^{\lhd d}_{\rhd e}\lambda \quad &\text{iff} & \Mmf, t' \models \lambda \text{ for all } t' \text{ such that }\\
 &&  t -t' \rhd e \text{ and } t - t' \lhd d,\\
\notag \Mmf, t \models \diamondplus^{\lhd d}_{\rhd e}\lambda \quad &\text{iff} & \Mmf, t' \models \lambda \text{ for some } t' \text{ such that }\\
 && t' -t \rhd e \text{ and } t' - t \lhd d,\\
\notag \Mmf, t \models \diamondminus^{\lhd d}_{\rhd e}\lambda \quad &\text{iff} & \Mmf, t' \models \lambda \text{ for some } t' \text{ such that }\\
 && t -t' \rhd e \text{ and } t - t' \lhd d. \label{prop-semantics4}
\end{align}
We say that $\Mmf$ satisfies a data assertion $P@\iota$ if $\Mmf, t \models P$ for \emph{all} $t \in \iota$. We say that $\Mmf$ satisfies an ontology axiom $\lambda \leftarrow \lambda_1 \land \dots \land \lambda_k$ (respectively, $\bot \leftarrow \lambda_1 \land \dots \land \lambda_k$), if $\Mmf, t \models \lambda_i$, for all $i = 1,\dots,k$, imply $\Mmf, t \models \lambda$ (resp., $\Mmf, t \models \bot$), for {\emph{every} $t \in \mathfrak{T}$.
Thus, the ontology axioms are \emph{global}. We say that $\Mmf$ satisfies a data instance $\D$ (resp., ontology $\O$) if it satisfies each statement in it. Finally, we say that $\Mmf$ satisfies a knowledge base $(\O, \D)$ and write $\Mmf \models (\O, \D)$ if $\Mmf$ satisfies both $\O$ and $\D$.

Our main reasoning problem is \emph{query answering}. Define an \emph{atomic query} (AQ) as an expression $P@\delta$, where $P$ is a proposition and $\delta$ is an \emph{interval variable}. An ontology $\O$ and an AQ $P@\delta$ constitute an \emph{ontology-mediated query} (OMQ) $\omq(\delta) = (\O,P@\delta)$. A \emph{certain answer} to $\omq(\delta)$ over $\D$ is any interval $\iota \in \int(\mathfrak{T})$ such that $\Mmf \models (\O, \D)$ implies $\Mmf, t \models P$ for all $t \in \iota$.

\paragraph{$\hMTL^\Box$ fragment.} We consider one important fragment $\hMTL^\Box$ of $\hMTL$, where the operators $\diamondplus^{\lhd d}_{\rhd e}$ and $\diamondminus^{\lhd d}_{\rhd e}$ are disallowed in the \emph{heads} of the rules. Note that each $\hMTL^\Box$ KB can be converted to KB that has $\boxplus^{\lhd d}_{\rhd e}$ and $\boxminus^{\lhd d}_{\rhd e}$ operators only, and the original KB is a conservative extension of it. For example, an axiom $R \leftarrow P \land \diamondminus^{\lhd d}_{\rhd e} Q$ can be replaced by the pair of axioms $R \leftarrow P \land Q'$ and $\boxplus^{\lhd d}_{\rhd e} Q' \leftarrow Q$. Finally, we consider a \emph{non-recursive} fragment $\hMTL^\Box_{nr}$ of $\hMTL^\Box$ by adopting the simplest definition of non-recursivivity: consider the relation $\prec$ on the symbols of $\O$ defined as $P \prec Q$ iff there is an axiom in $\O$, where $P$ occurs in the head and $Q$ in the body ($P$ \emph{depends on} $Q$). We require that $P \prec^* P$ for no symbol $P$ in $\O$, where $\prec^*$ is a transitive closure of $\prec$.

\paragraph{$\dMTL$.}\hspace*{-2mm} Consider the predicate symbols $P_0, P_1,
\dots$, each of some arity $m \geq 0$, and a set of object variables $x_0, x_1, \dots$. Data instances $\D$ here contain assertions $P(\avec{c})@\iota$, where $P$ is an $m$-ary predicate symbol, $\avec{c}$ an $m$-tuple of individual constants, and $\iota \in \int(\mathfrak{T})$. This assertion says that $P(\avec{c})$ is true at $\iota$. We denote by $\ind(\D)$ the set of all individual constants in $\D$. An ontology $\O$ is a finite set of axioms of the form~\eqref{clauses} with the literals $\lambda$ defined by the grammar:
$$\lambda \ ::= (\tau \neq \tau')\ \ \mid \ \ (\tau = \tau')\ \ \mid \ \ P(\avec{x}) \ \mid \ \ \mathbf{O}^{\lhd d}_{\rhd e} \lambda,$$
where $P$ is a predicate symbol of arity $m$, $\avec{x}$ is a vector of $m$ variables, and $\tau,\tau'$ are individual terms, i.e., variables or constants. We also impose other standard datalog  restrictions on our programs, and forbid (in)equality predicates in the heads. We call the predicates occurring in $\D$ \emph{extensional} and those occurring in the head of the axioms of $\O$ \emph{intentional}.
An \emph{interpretation}, $\Mmf$, is based on the domain $\Delta = \ind(\D)$ (for the individual variables and constants) and $\mathfrak T$. For any $m$-ary predicate $P$, $m$-tuple $\avec{c}$ from $\Delta$ and $t \in \mathfrak{T}$, $\Mmf$ specifies whether $P$ is \emph{true on $\avec{c}$ at $t$}, in which case we write $\Mmf,t \models P(\avec{c})$. Let $\nu$ be an \emph{assignment} of elements of $\Delta$ to individual terms (we adopt the standard name assumption: $\nu(c) = c$, for every individual constant $c$). We set:
\begin{align*}
 \Mmf, t \models^\nu \tau \ne \tau' \ &\text{iff } \  \nu(x_0) \ne \nu(x_1),\\
\Mmf, t \models^\nu \tau = \tau' \ &\text{iff } \  \nu(x_0) = \nu(x_1),\\
 \Mmf, t \models^\nu  P(\avec{x}) \ &\text{iff }\ \Mmf, t \models^\nu P(\nu(\avec{x})),
\end{align*}
and use inductively the formulas~\eqref{prop-semantics1}--\eqref{prop-semantics4}
with $\models^\nu$ instead of $\models$ for the cases $\mathbf{O}^{\lhd d}_{\rhd e} \lambda$. We say $\Mmf$ satisfies an ontology axiom $\lambda \leftarrow \lambda_1 \land \dots \land \lambda_k$ (respectively, $\bot \leftarrow \lambda_1 \land \dots \land \lambda_k$), if $\Mmf, t \models^\nu \lambda_i$ for each $i$ implies $\Mmf, t \models^\nu \lambda$ (resp., $\Mmf, t \models^\nu \bot$), for {\emph{every} $t \in \mathfrak{T}$ and assignment $\nu$. Finally, $\Mmf$ satisfies a data assertion $P(\avec{c})@\iota$ if $\Mmf, t \models P(\avec{c})$ for each $t \in \iota$, and $\Mmf \models (\O, \D)$ is defined straightforwardly.

  AQs are defined as $P(\avec{x})@\delta$, where $P$ is a predicate symbol of
  arity $m$, and $\delta$ is an interval
  variable. An ontology-mediated query is defined
  $\omq(\avec{x}, \delta) = (\O,P(\avec{x})@\delta)$. A \emph{certain
    answer} to $\omq(\avec{x},\delta)$ over $\D$ is any pair
  $(\avec{c}, \iota)$, such that $\avec{c} = \nu(\avec{x})$ for some
  $\nu$, and $\Mmf \models (\O, \D)$ implies
  $\Mmf, t \models P(\avec{c})$ for all $t \in \iota$.

Note that $\hMTL$ is a fragment of $\dMTL$ (where all predicates have arity $0$). We also consider the fragments $\dMTL^\Box$ and $\dMTL^\Box_{nr}$ defined with the same syntactic restrictions as $\hMTL^\Box$ and $\hMTL^\Box_{nr}$.

\section{Weather Use Case}

Our OBDA approach can be used to analyze meteorological data through ontology-mediated queries. The Meso\-West\footnote{\url{http://mesowest.utah.edu/}} project makes publicly available historical records of the weather stations across the US. This data is available in the relational tables {\sf Weather} containing the following fields:
\begin{description}
  \item[{\sf ID}.] Station ID. Example: KHYS.
  \item[{\sf TIME}.] Timestamp. Example: 11-11-2015 8:55 CST.
  \item[{\sf TMP}.] Temperature. Example: 15.6$^\circ$ C.
  \item[{\sf SKNT}.] Wind Speed. Example: 9.2 km/h.
  \item[{\sf P01I}.] Precipitation in 1 hour. Example: 0.09 cm.
\end{description}
Moreover, there are metadata tables {\sf Metadata} containing, in particular, location information of stations in the fields:
\begin{description}
  \item[{\sf ID}.] Station ID. Example: KHYS.
  \item[{\sf COUNTY.}] Example: Ellis.
  \item[{\sf STATE.}] Example: Kansas.
\end{description}
We can conceptualize this raw data by means of the SQL mappings. For
example, to extract the data for the extensional predicate
${\sf Precipitation}(x)@\langle t_1, t_2 \rangle$ (with the meaning
precipitation occurs at $x$ during $\langle t_1, t_2 \rangle$), we can use the following SQL query:
\begin{lstlisting}
  SELECT ID AS $x$,
    lag(TIME) over (partition
          by ID order by TIME) AS $t_1$,
    TIME AS $t_2$, "(" AS $\langle$, "]" AS $\rangle$
  FROM Weather
  WHERE P01I > lag(P01l)
    over(partition by ID order by TIME)
\end{lstlisting}
%
That is, we extract the intervals of the shape $(t_1, t_2]$, where $t_1$ and $t_2$ are the two \emph{next} timestamps for a given station. The ends of the interval are chosen to reflect the fact that, e.g., the precipitation is measured \emph{accumulatively} and the device produces the output in the end of the measurement interval. Analogously to ${\sf Precipitation}$, we populate by the data the other extensional predicates, such as ${\sf PositiveTemp}$ (temperature well above 0$^\circ$ C), ${\sf HurricaneForceWind}$ (wind with the speed above 118 km/h), ${\sf TempAbove24}$ and ${\sf TempAbove41}$ (temperature above 24 and 41$^\circ$ C).

Consider the ontology containing the axioms:
 \begin{align*}
     & {\sf Rain}(x)   \leftarrow   {\sf PositiveTemp}(x) \land  {\sf Precipitation}(x),\\
     \boxminus^{\leqslant 1h}_{\geqslant 0} & \mathsf{Hurricane}(x) \leftarrow  \boxminus^{\leqslant 1h}_{\geqslant 0} \mathsf{HurricaneForceWind}(x),\\
     \boxminus^{\leqslant 24 h}_{\geqslant 0} & \mathsf{ExcessiveHeat}(x) \leftarrow  \boxminus^{\leqslant 24 h}_{\geqslant 0} \mathsf{TempAbove24}(x) \land \\
      &\hspace{3.5cm} \diamondminus^{\leqslant 24 h}_{\geqslant 0} \mathsf{TempAbove41}(x),
 \end{align*}
The second axiom is already discussed in the introduction (here we use a slightly modified version to say that hurricane holds also at the time point, when the hurricane force wind begins), whereas the last axiom formalizes the definition of the situation when an excessive heat warning should be issued according to the US Weather Forecast Offices (24 hours with the minimal temperature above 24$^\circ$ C and the maximal above  41$^\circ$ C).

 We can also populate the binary
 predicate ${\sf LocationOf}(x, y)@\langle t_1, t_2 \rangle$ by using:
\begin{lstlisting}
 SELECT COUNTY AS $x$, ID AS $y$,
   $-\infty$ AS $t_1$, $\infty$ AS $t_2$, "(" AS $\langle$, ")" AS $\rangle$
 FROM Metadata
\end{lstlisting}
Note that we assume that ${\sf LocationOf}$ holds between a county and a station \emph{globally}. It is now possible to define:
\begin{align*}
  &{\sf HurricaneAffectedCounty}(x) \leftarrow\\
  & \hspace{3cm} {\sf LocationOf} (x,y) \land {\sf Hurricane}(y),\\
  &{\sf SpreadRainCounty}(x) \leftarrow {\sf LocationOf}(x,y) \land \\
  &\hspace{1cm}{\sf LocationOf}(x,z) \land
  (y \neq z) \land {\sf Rain}(y) \land {\sf Rain}(z).
\end{align*}

\section{Query Answering in $\dMTL^\Box_{nr}$}

In this section we first present an algorithm for computing certain
answers to an $\hMTL^\Box_{nr}$ OMQ $\omq(\delta) = (\O,P@\delta)$
over
$\D$. 

\paragraph{Normal form for $\hMTL^\Box_{nr}$.} Our procedure works on the ontology $\O$ containing only the clauses of the shape:
\begin{align*}
   P \leftarrow Q \land R, & \quad \bot \leftarrow Q \land R,\\
  \boxplus^{\lhd d}_{\rhd e} P \leftarrow Q, & \quad \boxminus^{\lhd d}_{\rhd e} P \leftarrow Q,\\
   P\leftarrow \boxplus^{\lhd d}_{\rhd e} Q, & \quad P \leftarrow \boxminus^{\lhd d}_{\rhd e} Q
\end{align*}
It is an easy exercise to verify that every  $\hMTL^\Box_{nr}$ can be brought to the normal form by performing the following operations:
\begin{itemize}
  \item[--] Substitute the axioms of the shape $\lambda \leftarrow \lambda_1 \land \dots \land \lambda_k$ for $k \geq 3$ by $k-1$ axioms with binary conjunctions using fresh symbols. Analogously for the axioms with $\bot$ in the head.
  \item[--] Remove $\Diamond^{\lhd d}_{\rhd e} \lambda$ literals in the body of the axioms as sketched in Preliminaries.
  \item[--] Remove the nested modalities $\Box^{\lhd d}_{\rhd e} \lambda$ by substituting them for $\Box^{\lhd d}_{\rhd e} P_\lambda$, for a fresh symbols $P_\lambda$, and adding:
      \begin{itemize}
       \item $P_\lambda \leftarrow \lambda$, if $\Box^{\lhd d}_{\rhd e} \lambda$ occurred in the body of the axiom,
       \item $\lambda \leftarrow P_\lambda$, if $\Box^{\lhd d}_{\rhd e} \lambda$ occurred in the head of the axiom.
      \end{itemize}
  \item[--] Remove the axioms of the shape $\lambda_0 \leftarrow \lambda_1 \land \lambda_2$, if $\lambda_i = \Box^{\lhd d}_{\rhd e} P$ for some $0 \leq i \leq 2$, as described in the previous step. Analogously for the axioms with $\bot$ in the head.
\end{itemize}
It can be readily verified that the resulting ontology in the normal form is in $\hMTL^\Box_{nr}$.

\paragraph{Algorithm.}
  We first assume that the facts of $\D$ are stored in the tables of the shape $P_i^*(t_1, t_2, \langle, \rangle)$, where $t_1, t_2 \in \mathfrak{T}$, $\langle$ is either $($ or $[$, and $\rangle$ is either $)$ or $]$. E.g., for $\D = \{P_i@(t_1, t_2], P_i@[t_1', t_2'] \}$ we produce the table $P_i^*$ with two tuples $\{\bigl(t_1, t_2, (, ]\bigr), \bigl(t_1', t_2', [, ]\bigr) \}$. Consider an intentional symbol $P$ and assume that for all $Q$ such that $P \prec Q$ the tables $Q^*$ are computed. Consider now the cases:

\smallskip
\noindent {$\bm{P \leftarrow Q \land R.}$} Then $P^*$ is computed as the \emph{minimal} table satisfying the condition:
      \begin{align*}
  Q^*\bigl(t_1, &t_2, \langle , \rangle\bigr)  \land  R^*\bigl(t_1', t_2', \langle' , \rangle'\bigr)  \land{} \\
  &{\sf ints}\bigl(t_1, t_2, \langle , \rangle, t_1', t_2', \langle' , \rangle'\bigr) \to
    P^*\bigl(t_1'', t_2'', \langle'', \rangle''\bigr),
\end{align*}
where ${\sf ints}(t_1, t_2, \langle , \rangle, t_1', t_2', \langle' , \rangle')$ is $\top$ if $\langle t_1, t_2 \rangle \cap \langle' t_1', t_2' \rangle' \neq \emptyset$ (the intervals intersect), otherwise it is $\bot$, and $\langle'' t_1'', t_2'' \rangle'' = \langle t_1, t_2 \rangle \cap \langle' t_1', t_2' \rangle'$ (the result of the  intersection). Note that $P^*$ is computed as a temporal join~\cite{Gao2005} of $Q^*$ and $R^*$. We also create a table $\bot^*$ for the axioms $\bot \leftarrow Q \land R$.

\smallskip
\noindent {$\bm{\boxplus^{\lhd d}_{\rhd e} P \leftarrow Q.}$} Then $P^*$ is computed as a minimal table satisfying:
\begin{align*}
  Q^*\bigl(t_1, t_2, \langle , \rangle\bigr) \to P^*\bigl(t_1+e, t_2+d, {\sf ed}\bigl(\langle, \rhd\bigr), {\sf ed}\bigl(\rangle, \lhd\bigr)\bigr)
\end{align*}
where the \emph{edge function} ${\sf ed}(\langle, \rhd)$ returns $[$, if $\langle$ is $[$ and $\rhd$ is $\geqslant$, and $($, otherwise. Then ${\sf ed}(\rangle, \lhd)$ is defined symmetrically. For example, if $Q^* = \{ \bigl(t_1, t_2, (, ]\bigr)\}$ and the axiom is $\boxplus^{< d}_{\geqslant e} P \leftarrow Q$, then $P^* = \{ \bigl(t_1+e, t_2+d, (, )\bigr)\}$. The axiom $\boxminus^{\lhd d}_{\rhd e} P \leftarrow Q$ is handled analogously.

\smallskip
\noindent $\bm{P \leftarrow \boxplus^{\lhd d}_{\rhd e} Q.}$ Consider the following example: let $Q^* = \{ \bigl( t_1, t_2, (, ]\bigr), \bigl( t_2, t_3, (, )\bigr)\}$ and the axiom $P \leftarrow \boxminus^{< d}_{\geqslant e} Q$ such that $d-e < t_3 - t_1$. Then, according to the semantics, $P^* = \{ \bigl( t_1-e, t_3-d, (, ]\bigr)\}$. In order to compute $P^*$ correctly we need to consider the \emph{concatenation} of the intervals $(t_1, t_2]$ and $(t_2, t_3)$. To compute $P^*$ in general we first produce a  \emph{closure} $Q'$ of $Q^*$ as the minimal table satisfying:
\begin{align*}
   Q^*\bigl(t_1,  t_2, \langle , \rangle\bigr) \to Q'\bigl(t_1,  t_2, \langle , \rangle\bigr),&\\
   Q^*\bigl(t_1, t_2, \langle , \rangle\bigr) \land Q'\bigl(t_1',  t_2', \langle' , \rangle'\bigr) & \land  (t_2' \leq t_2)  \land \\
  {\sf ints}\bigl(t_1, t_2, \langle , \rangle, t_1', t_2',& \langle' , \rangle'\bigr)   \to
    Q'\bigl(t_1', t_2, \langle', \rangle\bigr).
\end{align*}
After that $P^*$ can be obtained by:
      \begin{align*}
  Q'\bigl(t_1, t_2, \langle , \rangle\bigr) \land &~ {\sf fit}\bigl(t_1, t_2, \langle , \rangle, e, d, \rhd, \lhd \bigr) \to \\
   & P^*\bigl(t_1 - e, t_2 - d, {\sf de}\bigl(\langle , \rhd\bigr), {\sf de}\bigl(\rangle, \lhd\bigr) \bigr),
\end{align*}
where ${\sf fit}\bigl(t_1, t_2, \langle , \rangle, e, d, \rhd, \lhd \bigr)$ is $\top$, if there exists $t \in \mathfrak{T}$ such that $\{t + t' \mid t' \rhd e\text{ and }t' \lhd d \} \subseteq \langle t_1, t_2 \rangle$, and $\bot$ otherwise. Essentially, ${\sf fit}$ holds if the segment $\{t' \mid t' \rhd e\text{ and }t' \lhd d \}$ can be shifted so that it \emph{fits} inside $\langle t_1, t_2 \rangle$. Finally, another edge function ${\sf de}$ is needed to compute the ends of the resulting interval. Here ${\sf de}\bigl(\langle , \rhd\bigr)$ is $[$, if either $\langle$ is $($ and $\rhd$ is $>$, or $\langle$ is $[$ and $\rhd$ is $\geqslant$; otherwise ${\sf de}\bigl(\langle , \rhd\bigr)$ is $($. The definition of ${\sf de}\bigl(\rangle, \lhd\bigr)$ is symmetric. The axiom $\boxminus^{\lhd d}_{\rhd e} P \leftarrow Q$ is handled analogously. Observe that the computation of $Q'$ requires \emph{recursion}.

\smallskip
Clearly, when $P$ occurs in the head of several axioms, the table $P^*$ is taken equal to the union of the tables computed above. In fact, for every symbol $P$ in $\O$ the algorithm computes $P^*$ that, for a consistent KB $(\O, \D)$, satisfies:
\begin{itemize}
\item for every $t \in \mathfrak{T}$, there exists a certain answer $\iota$ to OMQ $\omq(\delta) = (\O,P@\delta)$ over $\D$ such that $t \in \iota$ iff there exists a tuple $\bigl( t_1, t_2, \langle, \rangle \bigr)$ in $P^*$ such that $t \in \langle t_1, t_2 \rangle$.
\end{itemize}
This correctness follows directly from the semantics of $\hMTL^\Box_{nr}$. Then, if the table $\bot^*$ is empty, as an output of the OMQ $\omq(\delta) = (\O,G@\delta)$ over $\D$ we produce the table $G^*$ (otherwise, we return $G^*$ with one special tuple $\bigl( -\infty, \infty, (,) \bigr)$ as $(\O, \D)$ is inconsistent). Clearly, the correctness above guarantees that $G^*$ represents the set of all certain answers.

One can extend the approach presented above to OMQ answering in
$\dMTL^\Box_{nr}$. Indeed, it is possible to convert an arbitrary
$\dMTL^\Box_{nr}$ ontology to the one in the normal form
\emph{similar} to that used above. The tables $P^*$ need to contain
the tuples of the shape
$\bigl( c_1, \dots, c_m, t_1, t_2, \langle, \rangle\bigr)$, where $m$
is the arity of $P$. The rules for processing the temporal axioms
essentially remain the same. The rules for computing the conjunctions
(joins) need to be adjusted to correctly handle the individual
arguments of the predicates.

\section{Discussion and Future Work}


\paragraph{Initial Experiments.} We made experiments to evaluate the
performance of the proposed algorithm on the
$\mathsf{Hurricane}(x)@\delta$ and $\mathsf{ExcessiveHeat}(x)@\delta$
OMQs with the ontology from the weather use case. We implemented the
algorithm of the previous section, for a given OMQ, as an SQL query
using {\sf WITH} clause and the {\sf RECURSIVE}~operator. That is, the
intermediate tables of the algorithm are defined as a sequence of
virtual SQL tables. The configuration of the computer that was used
for the experiments is Intel Core i5 @ 2.7 GHz, 8 GB RAM with 1867 MHz
DDR3 and OS X El Capitan operating system in version 10.11.4. The
weather data is stored in 64 bit PostgreSQL version 9.4.5. We ran
the queries over a table including 140\,881 rows. It took 3\,199 ms for
$\mathsf{Hurricane}$ and 481\,876 ms for $\mathsf{ExcessiveHeat}$ to
retrieve the results. We interpret this outcome as a positive
indication of the feasibility of our approach: even a straightforward
implementation appears to work. We foresee the following three
directions of the future work:

\paragraph{New Use Cases.} Our language is capable of expressing complex patterns of events that are of interest for such purposes as diagnostics of engines or devices. The axiom
\begin{multline*}
{\sf SmoothShutDown} \leftarrow {\sf IdleRPM} \land  \boxminus^{< 15{\it min}}_{>0} {\sf IntermRPM} \land{} \quad\mbox{}  \\
\mbox{} \hspace*{3cm} \diamondminus^{\leqslant 25min}_{\geqslant 15min} {\sf RunningRPM},
\end{multline*}
for instance, describes the event of smooth shutdown of an engine as being in an idle state after having intermediate speed (RpM) for 15 minutes and having a running speed before that (not further than 25 minutes). The axiom:
\begin{align*}
   {\sf ConsHighVibration} \leftarrow \boxminus^{\leqslant 50 sec}_{\geqslant 0} \diamondminus^{\leqslant 10 sec}_{>0} {\sf HighVibration}
\end{align*}
describes consistent high vibration as high vibration occurring every 10 seconds during a minute. Our OBDA approach seems to be able to capture many industrial use-cases. In the future, we plan to investigate such potential applications.

\paragraph{Open Theoretical Problems.}

At the moment, we do not know whether OMQ answering in $\hMTL$ is decidable. In fact, this question is open even for the fragment $\hMTL^\Box$. We plan to obtain complexity results for those languages, and we are particularly interested in \emph{data complexity} (that is, the complexity in the size of $\D$ when $\omq(\delta)$ is assumed to be fixed). It is also important to understand how the complexity results for $\hMTL$ carry over to $\dMTL$. To achieve our goal, we plan to study various techniques developed in the area of metric temporal logics~\cite{Ouaknine:2005:DMT:1078035.1079694,Ouaknine:2008:RRM:1432153.1432155,Hirshfeld2005148}
and modal logics over metric spaces~\cite{DBLP:journals/tocl/KutzWSSZ03,DBLP:journals/apal/SheremetWZ10,DBLP:journals/jsyml/WolterZ05}.

\paragraph{Implementation and Optimizations.}

The proposed query answering algorithm for $\dMTL^\Box_{nr}$ clearly allows for optimizations. For example, computing the transitive closure of the table $Q^*$ when processing the axiom $P \leftarrow \boxplus^{\lhd d}_{\rhd e} Q$ seems to be avoidable. Moreover, our algorithm does not make any assumption regarding the temporal ordering of the tuples. If such a realistic assumption is made, we may be able to develop more efficient algorithms, in particular, by using indexes on timestamps.

\bibliographystyle{aaai}
\bibliography{ontolp16}

\end{document}